\documentclass[12pt]{article}
\usepackage[xdvi]{graphicx}
\usepackage{amssymb}
\newcommand{\bea}{\begin{eqnarray}}
\newcommand{\eea}{\end{eqnarray}}

\makeatletter
\@addtoreset{equation}{section}
\makeatother

\def\ignore#1{{}}

\begin{document}
\begin{titlepage}
\begin{flushright}
OU-HET 665/2010
\end{flushright}

\vspace{25ex}

\begin{center}
{\Large\bf 
Asymptotic behavior of   
Lorentz violation on orbifolds
}
\end{center}

\vspace{1ex}

\begin{center}
{\large
Nobuhiro Uekusa
}
\end{center}
\begin{center}
{\it Department of Physics, 
Osaka University \\
Toyonaka, Osaka 560-0043
Japan} \\
\textit{E-mail}: uekusa@het.phys.sci.osaka-u.ac.jp
\end{center}


\vspace{3ex}

\begin{abstract}

Momentum dependence of quantum corrections with
higher-dimensional Lorentz
violation is examined
in electrodynamics on orbifolds.
It is shown that effects of the Lorentz violation 
are not decoupled at high energy scales.
Despite the 
loss of the higher-dimensional Lorentz invariance,  
a higher-dimensional 
Ward identity is found to be fulfilled for one-loop 
vacuum polarization.
This implies that gauge invariance may be prior to
Lorentz invariance as a guiding principle in
higher-dimensional field theory.
As a universal application of electrodynamics,
an extra-dimensional aspect for
Furry's theorem is emphasized.

\end{abstract}
\end{titlepage}


\section{Introduction \label{sec:intro}}

Field theory with extra dimensions provides an
interesting framework for physics beyond the 
standard model.
As in the four-dimensional case,
one of the fundamental keys that characterize
theory is symmetry which is 
preserved or broken.
In models with extra dimensions, a 
variety of symmetry breaking have been 
provided~\cite{Scherk:1978ta}-\cite{Haba:2009iv}.
It has also been shown that
combinations of sources for
extra-dimensional symmetry breaking
are relatively accommodating 
and yield various possibilities~\cite{Uekusa:2007sw}-%
\cite{Uekusa:2008bd}.
Associated with non-renormalizable properties,
it is still controversial whether
quantum corrections are validly extracted
in the field-theoretical context.
Higher-dimensional field 
theory can be regarded as a high-energy
effective theory with a distinct
ultraviolet completion.
While attempts for realistic models have
been developed,
most of models 
such as orbifold models
with a minimal setup
require higher-dimensional Lorentz invariance
as a basic symmetry.
However, extra dimensions
are clearly different than our four dimensions.
It includes potentially an
extra-dimensional Lorentz violation.

If the Lorentz invariance in extra dimensions 
is violated,
it is important to be taken into account whether
the symmetry breaking 
is spontaneous or not.
When a
symmetry breaking is described to be spontaneous
in a certain theory,
the corresponding 
symmetry is expected to be recovered at high energies
in its framework.
The Higgs mechanism 
is this type 
of symmetry breaking.
Symmetry breaking in orbifolding involves
the extra-dimensional origin.
It is nontrivial whether symmetry is recovered 
at high energy scales.
Even if the starting action is
Lorentz invariant,
loop effects can give rise to a Lorentz violation.
If a model in the standpoint of effective field theory 
beyond the standard model
allows that
the Lorentz invariance is lost
at high energy scales, 
the starting action
should be described in a Lorentz-non-invariant manner
or only
approximately in a Lorentz-invariant manner
with respect to extra dimensions.  
The extra-dimensional Lorentz violation 
has been found to affect
spectra, Kaluza-Klein parity and parity 
violation~\cite{Rizzo:2005um}.
Therefore in the field-theoretical context
it should be clarified if
the extra-dimensional Lorentz invariance
on orbifold models is asymptotically preserved.

In this paper, we study 
momentum dependence of
Lorentz violating terms 
in electrodynamics on an orbifold
$S^1/Z_2$.
With an explicit analysis for loop diagrams and 
renormalization,
it is shown that effects of the Lorentz violation 
are not decoupled at high energy scales.
As another notable feature, despite the 
loss of the higher-dimensional Lorentz invariance,  
a higher-dimensional 
Ward identity is found to be fulfilled for one-loop 
vacuum polarization.
This implies that higher-dimensional 
gauge invariance may be prior to
higher-dimensional 
Lorentz invariance as a guiding principle in
a high-energy field theory.
We also discuss
an extra-dimensional aspect for
Furry's theorem.

The paper is organized as follows.
In Sec.~\ref{sec:model},
our Lorentz violent action is given.
In Sec.~\ref{sec:renorm}, a formalism of
a renormalization is shown in the orbifold model.
In Sec.~\ref{sec:asymp},
the asymptotic energy dependence of Lorentz violating
terms is given.
It is also shown that
higher-dimensional Ward identity is
fulfilled for one-loop vacuum polarization.
In Sec.~\ref{sec:furry},
a discussion about Furry's theorem
is given.
In Sec.~\ref{sec:concl},
we conclude with some remark.
The detail of loop corrections is summarized in 
Appendix~\ref{ap:loop}.

\section{Five-dimensional electrodynamics
and Lorentz violation 
\label{sec:model}}

We start with the action for 
five-dimensional quantum electrodynamics,
\bea
  S = S_{LI} + S_{L V} + S_{GF} ,
   \label{fulla}
\eea
with the Lorentz invariant action,
\bea
  S_{LI} = \int d^4 x \cdot {1\over 2}
    \int_{-L}^L
      dy 
      \left(
      -{1\over 4} F_{MN} F^{MN}
      +  \bar{\psi} i\gamma^M D_M \psi \right) ,
      \label{LIa}
\eea
and the Lorentz violating action
\bea
    S_{L V} =
     \int d^4x \cdot {1\over 2}
     \int_{-L}^L
     dy
     \left(  -{\lambda \over 2} F_{\mu y} F^{\mu y} 
     + k\bar{\psi} i\gamma^5 D_y \psi\right) ,
     \label{LVa}
\eea
where $\lambda$ and $k$ are dimensionless coupling
constants and their nonzero values indicate the
violation of
the five-dimensional Lorentz invariance.
After a renormalization, both of 
$\lambda$ and $k$ are momentum-dependent.
The Lorentz violating terms such as
$\bar{\psi}\gamma_5 \psi$ can be 
absorbed by the terms in Eq.~(\ref{LVa})
via a field redefinition~\cite{Rizzo:2005um}.
The actions (\ref{LIa}) and (\ref{LVa}) 
have gauge invariance although its form
is not in a Lorentz-invariant way. 
The gauge fixing action is denoted as $S_{GF}$,
whose explicit form will be given after 
a field redefinition with respect to renormalization
factors.
The fifth-dimensional Lorentz violation
is only taken into account while
the four-dimensional Lorentz invariance is preserved.
The five-dimensional indices are denoted as $M$. 
Greek indices $\mu$ run over 0,1,2,3 and 
the fifth index is denoted as $y$.
The gamma matrices are given by
\bea
   \gamma^\mu =
     \left(\begin{array}{cc}
       & \sigma^\mu \\
       \bar{\sigma}^\mu & \\
       \end{array}\right) 
        , \qquad
     \gamma^5 =
     \left(\begin{array}{cc}
        -i {\bf 1}_2 & \\
        & i{\bf 1}_2 \\
        \end{array}\right) ,
\eea
where the Pauli sigma matrices are used as
$\sigma^\mu =
({\bf 1}_2, \sigma^i)$ 
and
$\bar{\sigma}^\mu =
(-{\bf 1}_2, \sigma^i)$.
The five-dimensional covariant derivative
is defined as
$D_M = \partial_M - ig A_M$.
The extra-dimensional space is compactified
on $S^1/Z_2$, where the fundamental region is
$0\leq y \leq L$.
The five-dimensional spacetime is flat
with the metric $(1, -1 , -1 , -1 , -1)$.
The orbifold boundary conditions
for gauge fields and fermions are
\bea
   A_\mu (x,-y) &\!\!\!=\!\!\!&
  A_\mu (x,y) ,\qquad
   A_\mu (x, L-y) = A_\mu (x,L+y) ,
\\
  A_y (x,-y) &\!\!\!=\!\!\!&
   -A_y (x,y) , \qquad
  A_y (x, L-y) = -A_y (x,L+y) ,
\\
  \psi (x, -y) &\!\!\!=\!\!\!&
  i\gamma^5
   \psi( x,y) , 
   \qquad
   \psi (x, L-y) =
   i\gamma^5 \psi (x, L+y) ,
\eea
such that the photon and left-handed Weyl fermion have 
zero mode.

In order to perform renormalized perturbation,
we define renormalized fields as
\bea
   A_\mu = Z_A^{1/2} A_{\mu r} , \qquad
   A_y = Z_5^{1/2} A_{y r} , \qquad
   \psi = Z_\psi^{1/2} \psi_r .
   \label{rfield}
\eea 
The Lagrangian terms for the gauge field
are rewritten as
\bea
   && 
 -{1\over 4} F_{MN} F^{MN} 
  -{\lambda \over 2} F_{\mu y} F^{\mu y}
\nonumber
\\
 &\!\!\!=\!\!\!&
-{1\over 4} F_{MN\, r} F^{MN}_r 
 -{\lambda_r \over 2} F_{\mu y \, r} F^{\mu y}_r 
\nonumber
\\
  && 
   -{1\over 4} \delta_1 F_{\mu\nu\, r} F_r^{\mu\nu}
   +{1\over 2} \delta_2 \partial_\mu A_{yr} \partial^\mu A_{yr}
   -\delta_4 \partial_\mu A_{yr} \partial_y A_{r}^\mu
 +{1\over 2} \delta_3 \partial_y A_{\mu r} \partial_y A_r^\mu ,
   \label{eqr}
\eea
where $\lambda_r$ is the renormalized coupling for
$\lambda$.
Among the counterterms
in the equation (\ref{eqr}),
the cross term 
$\partial_\mu A_{yr} \partial_y A_{r}^\mu$
also appears.
The renormalization factors are given by 
\bea
 && \delta_1 = Z_A -1 , \qquad
   \delta_2 =(1+\lambda)Z_5 -(1+\lambda_r) ,
\\
  &&
   \delta_3= (1+\lambda)Z_A -(1+\lambda_r) ,
  \qquad
   \delta_4 =(1+\lambda) Z_5^{1/2}
     Z_A^{1/2} -(1+\lambda_r) .
\eea
The part of the gauge field has
the original three coefficients
$\lambda$, $Z_A$ and $Z_5$.
One of the four
renormalization factors $\delta_1,\cdots, \delta_4$
can be written in terms of the other factors.
For example, $\delta_4$ is 
\bea
  \delta_4
    = (1+\lambda)(1+\delta_1)
      \left\{
      \left[
        {\delta_2 -\delta_3 \over
          (1+\lambda)( 1+\delta_1 )} +1\right]^{1/2}
          -1\right\}
          +\delta_3 .
\eea
The equation (\ref{eqr}) has gauge invariance
although it is not the 
five-dimensional Lorentz invariant form. 
It is convenient to choose
the gauge fixing action as
\bea
   S_{GF}
    = \int d^4x \cdot {1\over 2}
      \int_{-L}^L dy
      \left(
        -{1\over 2\xi}
          \left(\partial_\mu A_r^\mu
          -\xi (1+\lambda_r) 
  \partial_y A_{y r} \right)^2
          \right) .
\eea
For the gauge $\xi=1$, the kinetic term
and $\lambda_r$ term in Eq.~(\ref{eqr}) and
the gauge fixing yield
\bea
    -{1\over 2} \left[
      \partial_\mu A_{\nu r}
      \partial^\mu A_r^\nu
       -(1+\lambda_r) \partial_y A_{\mu r}
       \partial_y A_r^\mu \right]
   +{1\over 2} \left[
  \partial_\mu
   \tilde{A}_{y r}\partial^\mu
   \tilde{A}_{y r}
     - (1+\lambda_r)
      \partial_y \tilde{A}_{y r}
     \partial_y \tilde{A}_{y r} \right] ,
\eea
where the rescaling has been employed as
$\tilde{A}_{y r} \equiv \sqrt{1+\lambda_r} \,A_{y r}$
for the canonical normalization.
Unless $1+\lambda >0$, tachyonic degrees arise.
At the moment its positivity is assumed.
The cross terms of 
$A_\mu$ and $A_y$ are gathered into a total derivative
$-(1+\lambda_r) \partial_y 
(A_r^\mu \partial_\mu A_{y r})$,
which is vanishing due to periodicity.
From Eqs.~(\ref{fulla}) and (\ref{rfield}),
the Lagrangian terms for the fermion are
rewritten as
\bea
   \bar{\psi}_r i\gamma^M \partial_M \psi_r
   + k_r \bar{\psi}_r i\gamma^5 \partial_y \psi_r
   +\delta_5 \bar{\psi}_r i\gamma^\mu \partial_\mu \psi_r
  +\delta_6 \bar{\psi}_r i\gamma^5 \partial_y \psi_r ,
\eea
where $k_r$ is the renormalized coupling for $k$.
Correspondingly to the two coefficients
$k$ and $Z_\psi$,
the renormalization factors are given by
$\delta_5 = Z_\psi -1$ and
$\delta_6 =(1+k) Z_\psi -(1+k_r)$.
The Lagrangian terms of interactions are
rewritten as
\bea
   g_r A_{\mu r} \bar{\psi}_r \gamma^\mu  \psi_r
  + i{\cal N}_r
    g_r \tilde{A}_{yr}
      \bar{\psi}_r \gamma^5  \psi_r 
 + \delta_7
    A_{\mu r} \bar{\psi}_r \gamma^\mu  \psi_r
  +\delta_8
    \tilde{A}_{y r}
      \bar{\psi}_r \gamma^5  \psi_r  ,
\eea  
with the rescaled field $\tilde{A}_{yr}$
for $A_{y r}$. Here
${\cal N}_r \equiv -i(1+k_r)/\sqrt{1+\lambda_r}$.
The renormalization factors are
$\delta_7=
g Z_A^{1/2} Z_\psi -g_r$
and $\delta_8=\delta_8(\delta_1, \cdots , \delta_7,
k_r, \lambda_r , g_r)$.
For couplings and fields, the subscript $r$ and tilde 
to indicate renormalized and rescaled quantities 
will be suppressed hereafter.

In order to calculate
quantum loop corrections, we 
write down the four-dimensional
Lagrangian based on a mode expansion.
From the equations of motion, 
the mode expansion of fields is given by 
\bea
   A_\mu (x,y)
  &\!\!\!=\!\!\!&
   {1\over \sqrt{L}} A_{\mu 0} (x)
   +\sum_{n=1}^\infty
     \sqrt{2\over L}
    A_{\mu n} (x) \cos
   \left({n\pi \over L}y\right) ,
\\
   A_y (x,y)
   &\!\!\!=\!\!\!&
   \sum_{n=1}^\infty
   \sqrt{2\over L} A_{y n} (x)
  \sin \left( {n\pi \over L}y\right) ,
\\
   \psi_L (x,y) 
     &\!\!\!=\!\!\!&
       {1\over \sqrt{L}}
         \psi_{L 0} (x)
      +\sum_{n=1}^\infty
        \sqrt{2\over L}
           \psi_{Ln}(x) \cos
            \left({n\pi \over L}y\right) ,
\\
  \psi_R (x,y)
    &\!\!\!=\!\!\!&
      \sum_{n=1}^\infty
     \sqrt{2\over L}
       \psi_{Rn}(x) \sin \left(
         {n\pi \over L}y\right) .
\eea
After the integration of the fifth space,
the four-dimensional Lagrangian is obtained as
\bea
  {\cal L}_{4D} 
    = {\cal L}_{A_\mu}^{\textrm{\scriptsize quad}}
     +{\cal L}_{A_y}^{\textrm{\scriptsize quad}}
     +{\cal L}_{\textrm{\scriptsize cross}}^{%
       \textrm{\scriptsize quad}}
     +{\cal L}_{\psi}^{\textrm{\scriptsize quad}}
     +{\cal L}_{\textrm{\scriptsize int}} .
      \label{4dlag}
\eea
Here the quadratic Lagrangians are given by
\bea
   {\cal L}_{A_\mu}^{\textrm{\scriptsize quad}}
  &\!\!\!=\!\!\!&
   -{1\over 2} \partial_\mu A_{\nu 0} 
  \partial^\mu A_0^\nu
  -{1\over 2} \sum_{n=1}^\infty
  \left(
   \partial_\mu A_{\nu n} \partial^\mu A_n^\nu
    -m_{A n}^2 A_{\mu n}A_n^\mu\right)
\nonumber
\\
  &&
   -{1\over 4} \delta_1 F_{\mu\nu 0}F^{\mu\nu}_0
  -{1\over 2}
  \sum_{n=1}^\infty
  \left({1\over 2} \delta_1 F_{\mu\nu n}
   F_n^{\mu\nu}
   -{\delta_3 \over (1+\lambda)}
   m_{A n}^2 A_{\mu n}A_n^\mu \right) ,
    \label{laay0}
\\
   {\cal L}_{A_y}^{\textrm{\scriptsize quad}}
  &\!\!\!=\!\!\!&
  {1\over 2} \sum_{n=1}^\infty
  \left(\partial_\mu A_{y n} \partial^\mu A_{y n}
    -m_{A n}^2 A_{y n} A_{y n}\right)
   +
   {\delta_2 \over 2(1+\lambda)} \sum_{n=1}^\infty
    \partial_\mu A_{y n} \partial^\mu A_{y n}
  , 
   \label{laay}
\\
   {\cal L}_{\textrm{\scriptsize cross
  }}^{\textrm{\scriptsize quad}}
   &\!\!\!=\!\!\!& {\delta_4 \over (1+\lambda)}
      \sum_{n=1}^\infty m_{A n} 
  (\partial_\mu A_{y n}) A_{n}^\mu ,
  \label{laay2}
\\
  {\cal L}_{\psi}^{\textrm{\scriptsize quad}}
    &\!\!\!=\!\!\!&
      \bar{\psi}_0 i\gamma^\mu P_L \partial_\mu \psi_0
  +\sum_{n=1}^\infty \bar{\psi}_n 
  (i\gamma^\mu \partial_\mu -m_{\psi n}) \psi_n
\nonumber
\\
  &&
  +\delta_5 \bar{\psi}_0 i\gamma^\mu P_L \partial_\mu \psi_0 
  +\sum_{n=1}^\infty
   \bar{\psi}_n 
   \left(
    \delta_5 i\gamma^\mu \partial_\mu
   -{\delta_6 \over (1+k)} m_{\psi n}\right) \psi_n .
\eea
The Lagrangian 
${\cal L}_{A_\mu}^{\textrm{\scriptsize quad}}$
for $A_\mu$ has
counterterms for $\delta_1$ and $\delta_3$.
The Lagrangian 
${\cal L}_{A_y}^{\textrm{\scriptsize quad}}$
for $A_y$ has
a counterterm for $\delta_2$.
For the Lagrangian 
${\cal L}_{\textrm{\scriptsize cross
}}^{\textrm{\scriptsize quad}}$,
there is a cross term only for the counterterm.
The renormalization factor $\delta_4$ is
not independent of $\delta_1,\delta_2,\delta_3$.
The Lagrangian 
${\cal L}_{\psi}^{\textrm{\scriptsize quad}}$
for $\psi$ has
counterterms for $\delta_5$ and $\delta_6$.
The $n$-th masses of bosons and fermion are
\bea
  m_{A_\mu n} = \sqrt{1+\lambda}\, {n \pi \over L}
    =m_{A_y n} \equiv m_{A n},
\qquad
   m_{\psi n} =(1+k) {n\pi \over L} .
    \label{massn}
\eea     
We have defined Dirac fermions as
\bea
   \psi_0 \equiv
   \left( \begin{array}{c}
    \psi_{L 0} \\
     0 \\
    \end{array}\right) ,
  \qquad
  \psi_n \equiv
   \left( \begin{array}{c}
   \psi_{L n} \\
   \psi_{R n} \\
    \end{array}\right) ,
\eea
and introduced the left-chiral 
projection operator 
$P_L \equiv ({\bf 1}_2+i\gamma^5)/2$.
The interaction terms of the Lagrangian are
\bea
 && {\cal L}_{\textrm{\scriptsize int}}
  =
   {g\over \sqrt{L}}
       \bar{\psi}_0 \gamma^\mu P_L
       A_{\mu 0} \psi_0 
   +\sum_{n=1}^\infty {g\over \sqrt{L}}
      \bar{\psi}_n \gamma^\mu A_{\mu 0}
       \psi_n
\nonumber
\\
  &&
  +\sum_{n=1}^\infty
  {g\over \sqrt{L}}
  \left(\bar{\psi}_n \gamma^\mu
  P_L A_{\mu n} \psi_0
  +\bar{\psi}_0 \gamma^\mu P_L A_{\mu n} \psi_n \right)
\nonumber
\\
  &&
 + \sum_{n,m,\ell=1}^\infty 
  {g\over \sqrt{2L}}
  \left\{
   \bar{\psi}_n \gamma^\mu A_{\mu m}
   \psi_\ell
  \left(\delta_{n+m,\ell} + \delta_{n, m+\ell}\right)
 +\bar{\psi}_n \gamma^\mu i\gamma^5
   A_{\mu m} \psi_\ell \delta_{n+\ell, m} \right\}
\nonumber
\\
  && + \sum_{n=1}^\infty 
   {\cal N} 
  {g\over \sqrt{L}}
   \left(\bar{\psi}_n P_L A_{y n} \psi_0
   -\bar{\psi}_0 P_R A_{y n} \psi_n\right)
\nonumber
\\
 && + \sum_{n,m,\ell=1}^\infty 
    {\cal N}
   {g\over \sqrt{2L}}
   \left\{
   \bar{\psi}_n A_{y m} \psi_\ell
   \left(\delta_{n, m+\ell}
  -\delta_{n+m,\ell}\right)
    +\bar{\psi}_n i\gamma^5
 A_{y m} \psi_\ell \delta_{n+\ell, m} \right\} , 
   \label{lagint}       
\eea
where counterterms for interactions 
have been omitted. 
The sum of modes for three indices is denoted as
$\sum_{n,m,\ell=1}^\infty
=\sum_{n=1}^\infty 
\sum_{m=1}^\infty
\sum_{\ell=1}^\infty$.
At tree level, 
$\lambda$ and $k$ 
affect the Kaluza-Klein spectrum given 
in Eq.~(\ref{massn}).
The equation (\ref{laay}) means that
$A_{y n}$ has no counterterm for the mass.
As an explicit consistency check,
it will be shown that the one-loop
two-point function for $A_{y n}$
has the bulk divergence only for
a four-momentum term. 
In Eq.~(\ref{lagint}), the terms
$\bar{\psi}_n P_L A_{y n} \psi_0$
and $\bar{\psi}_0 P_R A_{y n} \psi_n$
have relative sign
and $\bar{\psi}_n A_{y m}\psi_\ell$
has the factor 
$(\delta_{n,m+\ell}-\delta_{n+m,\ell})$.
The importance of their signs will be emphasized 
in Sec.~\ref{sec:furry}.

\section{Renormalization on orbifolds
\label{sec:renorm}}

In this section, we give a formalism of
the renormalization
for two-point functions for $A_\mu$ and $A_y$.
The one-loop vacuum polarizations for $A_\mu$ and $A_y$
are diagonal with respect to Kaluza-Klein modes.
The detail of a calculation is summarized in
Appendix~\ref{ap:loop}.

\begin{figure}[htb]
\begin{center}
\vspace{2ex}

\includegraphics[width=12cm]{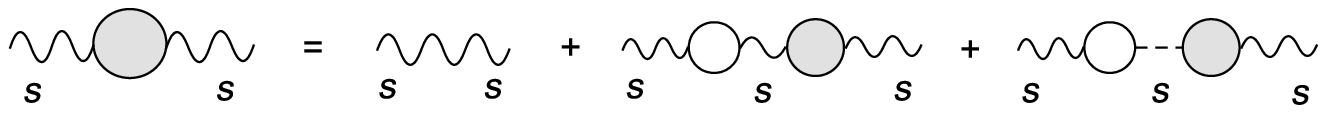}

\vspace{1ex}

\includegraphics[width=12cm]{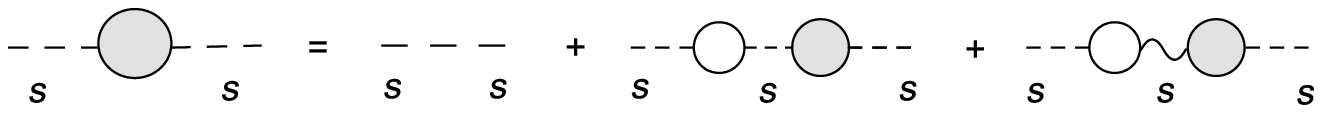}

\vspace{1ex}

\includegraphics[width=10cm]{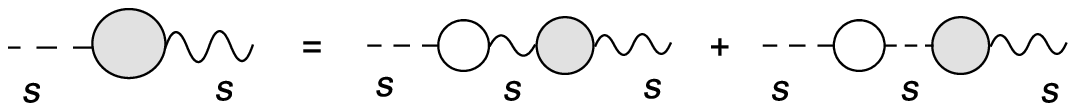}

\vspace{1ex}

\includegraphics[width=10cm]{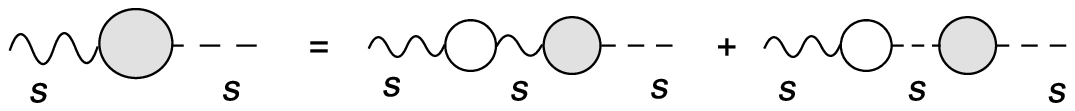}

\caption{One-loop decomposition
of exact propagators. \label{fig:prop}}
\end{center}
\end{figure}

The tree level propagators for the $s$-th fields
$A_{\mu s}$ and
$A_{y s}$ are
\bea
  D_{\mu\nu} (p^2) =
    {-i\eta_{\mu\nu}\over p^2 -m_{As}^2
  +i\epsilon} ,
\qquad
  D_{55} (p^2) =
    {i\over p^2 -m_{As}^2 +i\epsilon} ,
\eea
where $p^2 =p^\mu p_\mu$.
For simplicity, $i\epsilon$ will be omitted hereafter. 
Exact propagators can be decomposed with
one-particle irreducible amplitudes.
At one-loop level, diagrams of the decomposition are
shown in Figure~\ref{fig:prop},
where an unshaded circle denotes a one-loop diagram.
The corresponding equations are written as
\bea
   G_{\mu\nu} &\!\!\!=\!\!\!&
     D_{\mu\nu} + D_{\mu\rho} \Pi^{\rho \sigma}
       G_{\sigma \nu}
       +D_{\mu\rho} \Pi^{\rho 5} G_{5\nu} ,
\\
  G_{55} &\!\!\!=\!\!\!&
    D_{55}
      +D_{55} \Pi^{55} G_{55} 
      + D_{55} \Pi^{5\sigma} G_{\sigma 5} ,
\\
  G_{5\nu} &\!\!\!=\!\!\!&
     D_{55} \Pi^{5\sigma} G_{\sigma\nu}
       +D_{55} \Pi^{55} G_{5\nu} ,
\\
  G_{\mu 5} &\!\!\!=\!\!\!&
    D_{\mu\rho} \Pi^{\rho \sigma} G_{\sigma 5}
      +D_{\mu \rho} \Pi^{\rho 5} G_{55} .
\eea
The one-loop vacuum polarizations have the tensor 
structure given by
\bea
   \Pi_{\mu\nu} =
   \Pi_1 \eta_{\mu\nu} + \Pi_2 p_\mu p_\nu ,
\qquad
  \Pi_{\mu 5} =  -\Pi_\mu^{~\, 5} =
   \Pi_3 p_\mu 
  = \Pi_{5\mu} ,
\eea
where the explicit forms of 
$\Pi_1$, $\Pi_2$ and $\Pi_3$ will be given 
later. 
With these quantities,
the one-loop exact propagators
are solved as
\bea
 G_{\mu\nu} &\!\!\!=\!\!\!&
   -{D_{55}\over 1+D_{55}\Pi_1} \eta_{\mu\nu}
\nonumber
\\
    &&
   +{D_{55}\left[
   (D_{55}\Pi_2) (1-D_{55}\Pi^{55})
   +(D_{55}\Pi_3)^2\right] p_\mu p_\nu \over
   (1+D_{55}\Pi_1)
  \left[
  (1-D_{55}\Pi^{55}) 
  (1+D_{55}(\Pi_1 +\Pi_2 p^2))
   +(D_{55}\Pi_3)^2 p^2 \right] } ,
\\
   G_{55} &\!\!\!=\!\!\!&
  {D_{55} (1+D_{55} (\Pi_1 +\Pi_2 p^2))
  \over
  (1-D_{55}\Pi^{55})
  (1+D_{55}(\Pi_1 +\Pi_2 p^2))
   +(D_{55} \Pi_3)^2 p^2 } ,
\\
   G_{5\nu}&\!\!\!=\!\!\!&
   {D_{55}(D_{55}\Pi_3)
  \over 
  (1-D_{55}\Pi^{55})
  (1+D_{55}(\Pi_1 +\Pi_2 p^2))
   +(D_{55}\Pi_3)^2 p^2 } 
   p_\nu ,
\eea
where $G_{\nu 5} =G_{5\nu}$.

Now we perform the renormalization.
From the Lagrangians~(\ref{laay0}), (\ref{laay}) and 
(\ref{laay2}),
the contributions of counterterms are led to
\bea
   \Pi_{\mu\nu}^{ct} (p)
   &\!\!\!=\!\!\!&
    -i(p^2 \eta_{\mu\nu} -p_\mu p_\nu) \delta_1 
  + {i\delta_3 \over 1+\lambda} 
     m_{As}^2 \eta_{\mu\nu} ,
\\     
  \Pi_{\mu 5}^{ct} (p)
  &\!\!\!=\!\!\!&
  {\delta_4 \over 1+\lambda} m_{As}p_\mu ,
\qquad
  \Pi_{55}^{ct} (p) 
   = {i\delta_2 \over 1+\lambda} p^2 .
\eea
Only three renormalization factors among 
$\delta_1,\cdots,\delta_4$ are independent.
All the divergence associated with
$\Pi_1,\Pi_2,\Pi_3,\Pi^{55}$
must be removed with three renormalization factors.
As the first step,
it is convenient to fix the renormalization condition
for the off-diagonal component,
$G_{5\nu} (m_{As}^2) = 0$.
This condition yield 
\bea
  \Pi_3 (m_{As}^2) = 0 ,
   \label{condition1}
\eea
which corresponds to the fixing of 
$\delta_4$.
For $\Pi_3=0$, the other propagators are simplified as
\bea
   G_{\mu\nu} &\!\!\!=\!\!\!&
     -{D_{55}\over 1+D_{55}(\Pi_1 +\Pi_2 p^2)}
   \left[ \eta_{\mu\nu}
    + {D_{55}\Pi_2\over
    1+D_{55}\Pi_1} 
    (p^2 \eta_{\mu\nu} -p_\mu p_\nu)
    \right] ,
     \label{gmunu}
\\
  G_{55} &\!\!\!=\!\!\!&
   {D_{55}\over 1-D_{55}\Pi^{55}} . 
\eea
For Eq.~(\ref{gmunu}), $G_{\mu\nu}$,
the term of $(p^2 \eta_{\mu\nu} -p_\mu p_\nu)$
is renormalized with the counterterm for
$\delta_1$.
The corresponding 
renormalization condition can be imposed as
\bea
  \Pi_2(m_{As}^2) = 0 .
   \label{condition2}
\eea  
As we will show explicitly, the divergent part for 
$\Pi_1$ and $\Pi^{55}$ satisfy
$((\Pi_1+p^2\Pi_2)/m_{As}^2)_{\textrm{\scriptsize div}}
=(\Pi^{55}/p^2)_{\textrm{\scriptsize div}}$
at one-loop level.
This reduces to $\delta_2=\delta_3$.
Thus the renormalization can be chosen as
\bea
  \Pi_1(m_{As}^2) = 0 .
   \label{condition3}
\eea
On the other hand,
the finite part is
$((\Pi_1+p^2 \Pi_2)/m_{As}^2)
\not=(\Pi^{55}/p^2)$.
This means that the propagator for $A_y$ receives 
finite mass corrections
with $\Pi^{55}(m_{As}^2) \not= 0$.
For the divergent part,
it will be found in the following sections
that at one-loop level,
$\delta_2=\delta_3=\delta_4=(1+k)^2 \delta_1$.
Thus the momentum-dependent vacuum polarizations
$\Pi_1(p^2)$, $\Pi_2(p^2)$, $\Pi_3(p^2)$
and $\Pi^{55}(p^2)$ can be achieved
after the divergent part is fixed with
the renormalization
conditions (\ref{condition1}),
(\ref{condition2}) and (\ref{condition3}).
From these equations,
we can identify the asymptotic behavior
of the $\Pi_1(p^2)$, $\Pi_2(p^2)$, $\Pi_3(p^2)$
and $\Pi^{55}(p^2)$.
It needs to be checked
if Lorentz invariance is preserved
at high energy scales.

Renormalization for fermion self-energies
would be given in a similar procedure.
It may be technically complicated 
since one-loop self-energies are
not diagonal with respect to Kaluza-Klein modes.
This can be found from explicit one-loop amplitudes
summarized in Appendix~\ref{ap:loop}.  
A feasible way to treat off-diagonal components 
has been developed in Ref.~\cite{Uekusa:2010jf}.
At the first step to address asymptotic behavior
of the Lorentz violation, 
we are interest in
not only Lorentz invariance but also gauge invariance.
Both of these invariances can be simultaneously
examined when the vacuum polarization rather than
the self-energy is analyzed. 
Therefore we focus on the effects
on the vacuum polarization
for $A_\mu$ and $A_y$ and 
the issue for determining momentum-dependent
amplitudes with external fermions
will be left for future work.

\section{Energy dependence of Lorentz violating
terms and higher-dimensional Ward identity
\label{sec:asymp}}

Following the formalism of the previous section,
we analyze explicit one-loop results
for the Lorentz violation.
The one-loop contributions for the
vacuum polarization,
$\Pi_1^{(1)}$, $\Pi_2^{(1)}$, 
$\Pi_3^{(1)}$ and $\Pi^{55(1)}$ are
given via the dimensional regularization by
\bea
  \Pi_1^{(1)} (p^2)
  &\!\!\!=\!\!\!&
   {8 i g^2 \over (4\pi)^2 (1+k)} \int_0^1 dx 
   \left\{
     \left( z_4 - \sum_{n_p=1}^\infty
      z_3 e^{-{2 z_4\over z_3}} \cdot \cos 
  (2\pi n_p x s) \right)
 \right.
\nonumber
\\
   && \times  x(1-x)
     (p^2 -m_{\psi s}^2)
\nonumber
\\
  &&
  -\left. {1\over 4}
  \sum_{n_p=1}^\infty
     z_3  (z_3 + 2 z_4)
       e^{-{2 z_4\over z_3}}
    (1-2 x) m_{\psi s} \sin (2\pi n_p x s)
    \right\} ,
\\
  \Pi_2^{(1)} (p^2)
  &\!\!\!=\!\!\!&
  - {8 i g^2 \over (4\pi)^2 (1+k)} \int_0^1 dx 
   \left\{
     \left( z_4 - \sum_{n_p=1}^\infty
      z_3 e^{-{2 z_4\over z_3}} \cdot \cos 
  (2\pi n_p x s) \right)
 \right.
\nonumber
\\
   && \left.
   \times  x(1-x) 
\right\} ,
\\
  \Pi_3^{(1)} (p^2)
  &\!\!\!=\!\!\!&
   -{8 i g^2 {\cal N} \over (4\pi)^2 (1+k)}
    \int_0^1 d x
    \left\{
    \left( z_4 -\sum_{n_p=1}^\infty
  z_3 e^{-{2 z_4 \over z_3}} \cos (2\pi n_p x s) \right)
  \right.
\nonumber
\\
  && \times
     x(1-x) m_{\psi s}
\nonumber
\\
  && \left.
  +{1\over 4}  \sum_{n_p=1}^\infty
     z_3 (z_3 +2 z_4)
  e^{-{2 z_4 \over z_3}} (1-2 x) \sin (2\pi n_p xs)\right\} ,
\\
   \Pi^{55(1)} (p^2)
  &\!\!\!=\!\!\!&
  {8 i g^2 {\cal N}^2 
  \over (4\pi)^2 (1+k)}
   \int_0^1 d x
   \left\{
      z_4 x (1-x) p^2  \right.
\nonumber
\\
  &&
   -{1\over 4} \sum_{n_p=1}^\infty
     \left[
      3 z_3^2 (z_3 +2 z_4) + 2 z_3 (2 x (1-x)m_{\psi s}^2)
  \right]
   e^{-{2 z_4 \over z_3}} 
   \cos (2\pi n_p x s) 
\nonumber
\\
  && \left.
   +{1\over 4} \sum_{n_p=1}^\infty
      z_3 (z_3 +2 z_4) 
      e^{-{2 z_4 \over z_3}}
  (1-2 x) m_{\psi s}
    \sin (2\pi n_p x s) \right\} ,
\eea
where
$z_3 \equiv (1+k)/(n_p L)$ and
$z_4 \equiv \sqrt{x(1-x) (m_{\psi s}^2 -p^2)}$.
In the above equations,
the $n_p$-independent part is
finite due to the dimensional regularization in 
spacetime with odd dimensions
but it is
potentially divergent.
For the $n_p$-independent part,
$(\Pi_1^{(1)}+p^2 \Pi_2^{(1)})/m_{A s}^2
= \Pi^{55(1)}/p^2$ is satisfied.

From the renormalization conditions 
(\ref{condition1}),
(\ref{condition2}) and (\ref{condition3}),
the renormalization factors 
$\delta_1,\cdots , \delta_4$ are fixed.
Then the renormalized vacuum polarizations are
given by
\bea
  && \Pi_j (p^2)
    =
  \Pi_j^{(1)} (p^2)  -\Pi_j^{(1)} (m_{A s}^2) ,
\\
 && \Pi^{55}(p^2)
    =
    \Pi^{55(1)} (p^2)
\nonumber
\\
  &&  -ip^2
    \left[
         1-{i\Pi_1^{(1)} (m_{A s}^2)\over m_{A s}^2}
       -{1\over 1+i\Pi_2^{(1)} (m_{A s}^2)}
       \left(
       1-\Pi_3^{(1)} (m_{A s}^2)
       -{i\Pi_1^{(1)} (m_{A s}^2)\over m_{A s}^2}
   \right)^2
       \right] ,
\eea
where $j=1,2,3$.    
At high energies, 
the vacuum polarizations behave as
\bea
  \Pi^{\mu \nu} (p^2)&\!\!\!\to\!\!\!&
    \Pi^{\mu \nu}_{\textrm{\scriptsize as}} (p^2) =
    \left[ -(p^2 \eta^{\mu\nu} -p^\mu p^\nu)
     - {\cal N}^2 m_{A s}^2 \eta^{\mu\nu} \right]
      \Pi_2^{\textrm{\scriptsize as}} (p^2),
       \label{as1}
\\
  \Pi^{\mu 5}(p^2) &\!\!\!\to \!\!\!&
  \Pi^{\mu 5}_{\textrm{\scriptsize as}} (p^2) =
    -i{\cal N}^2 (p^\mu m_{A s}) 
   \Pi_2^{\textrm{\scriptsize as}} (p^2),
     \label{as2}
\\
  \Pi^{55}(p^2) &\!\!\!\to \!\!\!&
  \Pi^{55}_{\textrm{\scriptsize as}} (p^2) =
    -{\cal N}^2 p^2 \Pi_2^{\textrm{\scriptsize as}}(p^2)
     ,  \label{as3}
\eea
where 
$\Pi_2^{\textrm{\scriptsize as}}(p^2)
\equiv  
-8 ig^2(4\pi)^{-2} (1+k)^{-1}
\int_0^1 dx \, z_4 x(1-x)$.
In obtaining the asymptotic values~(\ref{as1}), 
(\ref{as2}) and (\ref{as3}), we have employed
the renormalization factors $Z_A$ and $Z_5$ and 
the renormalized coupling constant $\lambda$ 
so as to satisfy the renormalization conditions.
Explicitly these constants are given by
\bea
   Z_A &\!\!\!=\!\!\!&
     1+ i\Pi_2^{(1)}(m_{A s}^2) ,
\\
   Z_5 &\!\!\!=\!\!\!&
     {1\over 1+i\Pi_2^{(1)} (m_{A s}^2)}
     \left(1-\Pi_3^{(1)} (m_{A s}^2)
       -{i\Pi_1^{(1)}(m_{A s}^2)\over m_{A s}^2}
  \right)^2 ,
\\
   \lambda_r 
     &\!\!\!=\!\!\!&
    \lambda -(1+\lambda)
       {i\Pi_1^{(1)} (m_{A s}^2)\over m_{A s}^2} .
        \label{lamr}
\eea       
where the subscript $r$
indicates a renormalized quantity again
to avoid confusion.
The equations (\ref{as2}) and (\ref{as3})
include ${\cal N}_r$ in which $\lambda_r$ 
obeys Eq.~(\ref{lamr}) and 
is generally nonvanishing.
Thus the extra-dimensional
Lorentz invariance is violated 
in a generic region in the parameter space 
at high energy
scales.
Especially $\lambda =0$ does not mean
$\lambda_r=0$.    

To identify 
the effect of 
the violation of translation invariance 
due to the brane, we consider the limit $L\to \infty$.
For this limit,
the factor $z_3$ approaches zero, 
$z_3\to 0$ as $L^{-1}$ so that
the vacuum polarization
become Eqs.~(\ref{as1}), (\ref{as2}) and (\ref{as3}).
Then $\lambda_r$ is given in Eq.~(\ref{lamr}).
In the representation (\ref{lamr}), the limit yields
$\Pi_1^{(1)}(m_{A s}^2)\to 0$ as $L^{-3}$.
Thus the couping constant for $L\to \infty$ is
$\lambda_r \to \lambda$.
Therefore the infinite compactification radius
and zero original $\lambda$ can recover
the higher-dimensional Lorentz invariance.

Now we move on to the issue of 
Ward identity.
We compare the Lorentz
violating case with
a simple extension of the four-dimensional
quantum electrodynamics.
In a simple extension, 
the vacuum polarization has the form
$\Pi_{5D}^{MN} =(p^M p^N - p^L p_L \eta^{MN}) 
\Pi$.
This is decomposed as
\bea
 \Pi_{5D}^{\mu\nu} =
  \left[ 
  - \left(p^2 \eta^{\mu\nu} - p^\mu p^\nu
    \right) + p_5^2 \eta^{\mu\nu}
  \right]     
    \Pi ,
\qquad
  \Pi_{5D}^{\mu 5} = 
   (p^\mu p^5)\Pi ,
\qquad 
   \Pi_{5D}^{55} = 
   p^2 \, \Pi .
   \label{p5D}
\eea
which satisfy the identities,
\bea
   p_\mu \Pi_{5D}^{\mu\nu} +p_5 \Pi_{5D}^{5\nu} =0 , \qquad
   p_\mu \Pi_{5D}^{\mu 5} +p_5 \Pi_{5D}^{55}=0 .
\eea
On the other hand,
the asymptotic
vacuum polarization given in Eq.~(\ref{as1}),
(\ref{as2}) and (\ref{as3}) have the relation 
\bea
   p_\mu \Pi_{\textrm{\scriptsize as}}^{\mu\nu} 
  +i m_{A s} \Pi_{\textrm{\scriptsize as}}^{5\nu} =0 , 
\qquad
   p_\mu \Pi_{\textrm{\scriptsize as}}^{\mu 5} 
 -i m_{A s} \Pi_{\textrm{\scriptsize as}}^{55}=0 .
\eea
From the correspondence 
$\Pi_{\textrm{\scriptsize as}}^{\mu\nu}
\leftrightarrow \Pi_{5D}^{\mu\nu}$,
$\Pi_{\textrm{\scriptsize as}}^{\mu 5}
\leftrightarrow i\Pi_{5D}^{\mu 5}$
and
$\Pi_{\textrm{\scriptsize as}}^{55}
\leftrightarrow \Pi_{5D}^{55}$,
we find that the one-loop vacuum polarizations
satisfy the five-dimensional Ward identity
even without preserving the five-dimensional
Lorentz invariance.

\section{Furry's theorem on orbifolds
\label{sec:furry}}

So far we have examined the properties of
the vacuum polarizations with an explicit diagrammatic
calculation.
In this section, we give a formal aspect in 
higher-dimensional gauge theory. 

In the four-dimensional electrodynamics,
the charge conjugation is 
a symmetry of the theory,
$C|\Omega\rangle =|\Omega\rangle$, where
$C$ denotes the charge conjugation and
$|\Omega\rangle$ is the vacuum state.
The electromagnetic current, 
$j^\mu= \bar{\psi} \gamma^\mu\psi$ 
changes sign under the charge conjugation,
$Cj^\mu(x) C^\dag=-j^\mu(x)$
so that its vacuum expectation value is vanishing,
$\langle \Omega|T j^\mu(x)|\Omega\rangle
=0$.
Furry's theorem states
that any vacuum vacuum expectation value of
an odd number of electromagnetic currents
is vanishing.

Now we consider a two-current function 
${\cal M}^\mu_Y \equiv \langle T
j^\mu(x_1) j_Y(x_2) \rangle$ by introducing
another operator $j_Y=\bar{\psi}\psi$
and by imaging gauge and Yukawa interactions 
for external lines.
Here the ground state 
of the free theory with the symmetry of
the charge conjugation is denoted as $\rangle$.
Because of the charge conjugation
$C j_Y(x) C^\dag = + j_Y(x)$,
the two-current function ${\cal M}_Y^\mu$ is vanishing.
At the first sight,
the function ${\cal M}_Y^\mu$ 
with gauge and Yukawa interactions
seems to look like
the vacuum polarizations $\Pi^{\mu 5}$ and
$\Pi_{5D}^{\mu 5}$.
On the other hand, 
the vacuum polarization
$\Pi_{5D}^{\mu 5}$ is not vanishing 
for nonzero $\Pi_{5D}^{\mu\nu}$ as seen from
Eq.~(\ref{p5D}).
We have also explicitly derived a nonzero $\Pi^{\mu 5}$.
Thus the structure of $\Pi^{\mu 5}$ needs 
to be clarified 
from the viewpoint of Furry's theorem.

The one-loop two-point function
for $A_{\mu j}(x)$ and $A_{y s}(w)$ is given by
\bea
  && {\cal N} {g^2 \over 2L} \delta_{js}
    \int d^4 x_1 d^4 x_2
       D_{j ,\mu\rho} (x-x_1) D_s (w-x_2)
       \langle T{\cal O}_Y^\rho (x_1,x_2) \rangle ,
\eea
with the two-current operator
\bea
   {\cal O}_Y^\rho (x_1,x_2) 
  &\!\!\!\equiv \!\!\!& 
   j_{s,0}^\rho (x_1) 
   j_{0,s} (x_2) 
  - j_{0,s}^\rho (x_1) j_{s,0}(x_2)
\nonumber
\\
  && 
  - j_{n,\ell}^\rho (x_1)
  j_{\ell, n}(x_2)  
   (\delta_{n+s,\ell} -\delta_{n,s+\ell}
    )
   - j_{n,\ell}^{\rho 5} (x_1)
   j_{\ell, n}^5 (x_2) 
  \delta_{n+\ell, s}     
   . \label{oy}
\eea
Here the currents with zero mode are given by
$j_{I,J}^\rho =
\bar{\psi}_I \gamma^\rho \psi_J$ and
$j_{I,J} = \bar{\psi}_I \psi_J$,
where $I,J=0,1,\cdots,\infty$
and the currents with $\gamma^5$ are given by
$j_{n,\ell}^{\rho 5}= \bar{\psi}_n\gamma^\rho
i\gamma^5 \psi_\ell$ and
$j_{\ell,n}^5=\bar{\psi}_\ell i\gamma^5\psi_n$.
The charge conjugation yields
a change of the overall factor
and the interchange of indices,
\bea
 C j_{I,J}^{\rho}(x) C^\dag &\!\!\!=\!\!\!&
   -j_{J,I}^\rho (x) , 
 \qquad
 C j_{I,J}(x) C^\dag
   = + j_{J,I}(x) ,
\\
 C j_{n,\ell}^{\rho 5} (x) C^\dag
    &\!\!\!=\!\!\!&
    + j_{\ell, n}^{\rho 5}(x) ,
   \qquad
 C j_{n,\ell}^{5} (x) C^\dag
    = + j_{\ell, n}^{\rho 5}(x) .
\eea
From these equations, we
obtain 
$C{\cal O}_Y^\rho (x_1, x_2)C^\dag
= +{\cal O}_Y^\rho (x_1, x_2)$.
Therefore that $\Pi^{\mu 5}$ is not necessarily 
zero is consistent with Furry's theorem.  
In Eq.~(\ref{lagint}), 
$\bar{\psi}_n P_L A_{y n} \psi_0$
and $\bar{\psi}_0 P_R A_{y n} \psi_n$
have relative sign. If they have the 
same sign, the contribution
from the first line in Eq.~(\ref{oy})
would vanish.
The role of the relative sign in the term
$\bar{\psi}_n A_{y m}\psi_\ell$ 
in Eq.~(\ref{lagint})
is similar.

Application of Furry's theorem in
orbifold models may be 
given not only for two-point functions
but also for other functions.
For example,
the vacuum expectation value of one current is 
vanishing,
$\langle T j_{n,n}^\mu(x) \rangle=0$,
where the indices are the identical $n$.
Because a nonzero $\Pi^{\mu 5}$ is 
expected from a nonzero $\Pi_{5D}^{MN}$,
Furry's theorem 
in effective four-dimensional theory
may be related to the discrete symmetry in
the original higher-dimensional theory.
Due to the dependence of   
Lorentz transformation on
the dimensionality
of spacetime,
it is nontrivial 
to introduce discrete symmetry such as $P$, $C$, $T$
in higher-dimensional theory
\cite{Adachi:2009vy,Lim:2009pj}.
We leave further exploration of this issue 
for future work.

\section{Conclusion \label{sec:concl}}

We have studied 
the momentum dependence of
Lorentz violating terms in
the field-theoretical context
in electrodynamics on orbifolds.
Here an explicit analysis has been performed 
for loop diagrams and 
renormalization.
We have found that
the extra-dimensional Lorentz invariance is
violated in a generic region
in the parameter space at high energy scales.
In particular, even if the original action is
higher-dimensional Lorentz invariant,
it is violated by loop effects. 
While the higher-dimensional Lorentz invariance
is lost,  
a higher-dimensional 
Ward identity has been
 found to be fulfilled for the one-loop 
vacuum polarization.
Therefore higher-dimensional 
gauge invariance may be prior to
higher-dimensional 
Lorentz invariance as a guiding principle in
a high-energy field theory.
We have also discussed 
Furry's theorem in orbifold models to
confirm the consistency about 
the vacuum polarizations.

The four-dimensional Lorentz violation
has also been studied as a distinct topic
of Lorentz violation.
In the four-dimensional electrodynamics 
with Lorentz violation, it has been discussed that
Pauli-Villars regularization
is a useful choice associated with
gauge invariance~\cite{%
Jackiw:1999yp, PerezVictoria:2001ej, Altschul:2004gs}.
On the other hand, it has been shown that
propagators corresponding to 
Pauli-Villars are 
radiatively generated in an orbifold model~\cite{%
Uekusa:2009dy}.
In this light, the Pauli-Villars regulator may be
the necessity of an extra-dimensional 
model rather than a choice.
These relations should be examined further.

\vspace{8ex}

\subsubsection*{Acknowledgments}

This work is supported by Scientific Grants 
from the Ministry of Education
and Science, Grant No.~20244028.

\newpage

\begin{appendix}

\section{Loop corrections\label{ap:loop}}

In this appendix, the details of loop corrections 
are given.

\subsection{Diagrams and four-momentum integrals}

We evaluate loop corrections 
by calculating
the sum of diagrams for each Kaluza-Klein mode.
Propagators are defined for four-dimensional 
fields.
The tree-level propagators are diagonal with 
respect to Kaluza-Klein modes and are given by
\bea
   D^{\mu\nu} (x-w)
     &\!\!\!=\!\!\!&
     \langle T A_0^\mu (x) A_0^\nu (w) \rangle
    = \int {d^4 p\over (2\pi)^4}
     {-i\eta^{\mu\nu}\over
        p^2 +i\epsilon} e^{-ip\cdot
          (x-w)}
\\
   D_n^{\mu\nu} (x-w) &\!\!\!=\!\!\!&
   \langle T A_n^\mu (x) A_n^\nu (w) \rangle
  =  \int {d^4 p\over (2\pi)^4}
      {-i\eta^{\mu\nu}\over
       p^2 -m_{An}^2 +i\epsilon}
        e^{-ip\cdot (x-w)}
\\
  D_n (x-w) &\!\!\!=\!\!\!&
    \langle T A_{y n} (x) A_{y n} (w) \rangle
    \int {d^4 p\over (2\pi)^4}
      {i\over
       p^2 -m_{An}^2 +i\epsilon}
        e^{-ip\cdot (x-w)} ,
\eea
for bosons and
\bea
 S (x-w) &\!\!\!=\!\!\!&
  \langle T \psi_0(x) \bar{\psi}_0(w)\rangle
   = \int {d^4 p\over (2\pi)^4}
      P_L 
      {ip\!\!\!/\over p^2 +i\epsilon}
        e^{-ip\cdot (x-w)}
\\
  S_n (x-w) &\!\!\!=\!\!\!&
    \langle T \psi_n(x) \bar{\psi}_n (w)\rangle
   = \int {d^4 p\over (2\pi)^4}
  {i(p\!\!\!/ +m_{\psi n})\over
  p^2 - m_{\psi n}^2 +i\epsilon}
   e^{-i p\cdot (x-w)} ,  
\eea  
for fermions.

The vacuum polarizations for $A_\mu$ and $A_y$
involve the following momentum integrals:
\bea
    I_1 (\mu, \nu; m_1, m_2)
    &\!\!\!\equiv \!\!\!&
   -{g^2 \over 2L} 
    \int {d^4 p_1 \over (2\pi)^4}
    \textrm{tr} 
   \left(
   {p\!\!\!/{}_1 \over p_1^2 - m_1^2}
   \gamma_\mu 
  {p\!\!\!/{}_1 + p\!\!\!/{}_2 \over
   (p_1 +p_2)^2 -m_2^2}  \gamma_\nu \right)
    ,
\\
   I_2 (\mu , \nu; m_1 , m_2)
   &\!\!\!\equiv \!\!\!&
   -{g^2 \over 2L}
    \int {d^4 p_1 \over (2\pi)^4}
  \textrm{tr}\left(
   {m_1 \over p_1^2 -m_1^2}
   \gamma_{\mu} 
   {m_2\over (p_1+p_2)^2 -m_2^2}
  \gamma_\nu \right) ,
\eea
for two four-indices,
\bea
  I_1 (\mu; m_1, m_2) &\!\!\!\equiv \!\!\!&
   -{g^2 \over 2L} \int {d^4 p_1 \over (2\pi)^4}
 \textrm{tr}\left(
    {m_1 \over p_1^2 -m_1^2} \gamma_\mu
   {p\!\!\!/{}_1 + p\!\!\!/{}_2 
  \over
   (p_1 +p_2)^2 -m_2^2} \right) ,
\\
  I_2 (\mu; m_1, m_2)
   &\!\!\!\equiv \!\!\!&
    -{g^2\over 2L}
  \int {d^4 p_1 \over (2\pi)^4}
  \textrm{tr}
   \left( {p\!\!\!/{}_1 \over
    p_1^2 -m_1^2}
  \gamma_\mu
   {m_2 \over (p_1 + p_2)^2 -m_2^2} \right) ,
\eea
for one four-index and
\bea
  I_1 (m_1, m_2) &\!\!\!\equiv \!\!\!&
    {g^2 \over 2L}
      \int {d^4 p_1 \over (2\pi)^4}
        \textrm{tr}
         \left( {p\!\!\!/{}_1 \over
         p_1^2 -m_1^2}
           {p\!\!\!/{}_1 +p\!\!\!/{}_2
           \over
      (p_1 +p_2)^2 -m_2^2} \right)
   ,
\\
  I_2 (m_1 , m_2) &\!\!\!\equiv \!\!\!&
   {g^2 \over 2L} \int
   {d^4 p_1 \over (2\pi)^4}
  \textrm{tr}
  \left( {m_1\over p_1^2 -m_1^2}
  {m_2 \over (p_1 +p_2)^2 -m_2^2}\right) ,
\eea
for no four-indices.
These satisfy a property
$I_1 (\mu ; m_2 , m_1) = - I_2 (\mu ; m_1 , m_2)$.

The self-energies for $\psi$ involve the following
momentum integrals:
\bea
   B_1 (m_1, m_2) &\!\!\!\equiv \!\!\!&
      -{g^2 \over 2L}
        \int {d^4 p_2 \over (2\pi)^4}
         \gamma^\mu
         {1\over p_2^2 -m_1^2}
          {p\!\!\!/{}_1 -p\!\!\!/{}_2
          \over (p_1 -p_2)^2 -m_2^2}
          \gamma_\mu ,
\\
  B_2 (m_1 , m_2) &\!\!\!\equiv \!\!\!&
      -{g^2 \over 2L}
      \int {d^4 p_2 \over (2\pi)^4}
       \gamma^\mu
       {1\over p_2^2 -m_1^2}
       {m_2 \over (p_1-p_2)^2 -m_2^2} \gamma_\mu ,
  \\
   E_1 (m_1 , m_2) &\!\!\!\equiv \!\!\!&
     -{g^2 \over 2L}
       \int {d^4 p_2 \over (2\pi)^4}
        {1\over p_2^2 -m_1^2}
         {p\!\!\!/{}_1 -p\!\!\!/{}_2
         \over
         (p_1 -p_2)^2 -m_2^2}
         ,
\\
  E_2 (m_1, m_2)
   &\!\!\!\equiv \!\!\!&
    -{g^2 \over 2L}
      \int {d^4 p_2 \over (2\pi)^4}
       {1\over p_2^2 -m_1^2}
       {m_2 \over 
       (p_1-p_2)^2 -m_2^2} .
\eea

With these integral expressions,
the vacuum polarizations
are summarized as follows:
\bea
\begin{picture}(65,40)
    \put(0,0){\includegraphics[width=2cm]{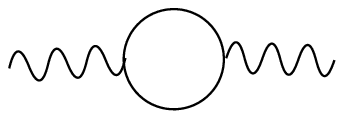}}
    \put(0,20){0}
    \put(50,20){0}
\end{picture}
  &=& 
     \sum_{n=-\infty}^\infty
       \left\{
         I_1 (\mu, \nu ; m_{\psi n} , m_{\psi n})
          + I_2 (\mu, \nu ; m_{\psi n} , m_{\psi n}) 
  \right\} ,
\\
\begin{picture}(65,40)
    \put(0,0){\includegraphics[width=2cm]{munu.eps}}
    \put(0,20){$j$}
    \put(50,20){$s$}
\end{picture}
  & = &
    \sum_{n=-\infty}^\infty
     \left\{
      I_1 (\mu,\nu; m_{\psi n}, m_{\psi , n+s})
      +I_2 (\mu,\nu; m_{\psi n} , m_{\psi , n+s})
      \right\}
      \delta_{j s} ,
\\
\begin{picture}(65,40)
    \put(0,0){\includegraphics[width=2cm]{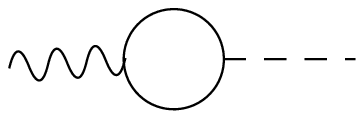}}
    \put(0,20){$j$}
    \put(50,20){$s$}
\end{picture}
 &=&  {\cal N}
   \sum_{n=-\infty}^\infty
     \left\{
     I_1 (\mu; m_{\psi n} ,m_{\psi, n+j})
     +I_2 (\mu ; m_{\psi n}, m_{\psi , n+j})
     \right\} \delta_{j s} ,
\\
\begin{picture}(65,40)
    \put(0,0){\includegraphics[width=2cm]{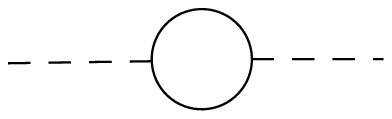}}
    \put(0,20){$j$}
    \put(55,20){$s$}
\end{picture}
 &=&  {\cal N}^2
   \sum_{n=-\infty}^\infty
     \left\{
       I_1 (m_{\psi n} , m_{\psi , n+j})
       +I_2 (m_{\psi n} , m_{\psi, n+j}) \right\} 
       \delta_{j s} .
\eea       
The vacuum polarizations for $A_\mu$ and $A_y$
do not give rise to one-loop corrections for 
brane terms.
The Kaluza-Klein modes for external lines
are diagonal.

The fermion self-energies are summarized as follows:
\bea
\begin{picture}(65,40)
    \put(0,0){\includegraphics[width=2cm]{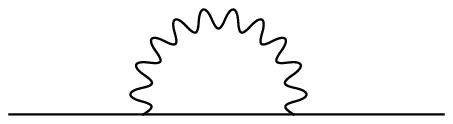}}
    \put(0,5){$0$}
    \put(50,5){$0$} 
\end{picture}
 &=& \sum_{n=-\infty}^\infty
    B_1 (m_{A n} , m_{\psi n}) 
    P_L
    + B_1 (0,0) P_L ,
\\
\begin{picture}(65,40)
    \put(0,0){\includegraphics[width=2cm]{sunrise.eps}}
    \put(0,5){$0$}
    \put(50,5){$s$} 
\end{picture}
&=&    \sqrt{2}
      \sum_{n=1}^\infty
       \left\{
  B_1 (m_{An}, m_{\psi n}) P_L
   + B_2 (m_{An}, m_{\psi n}) P_R 
  \right\} \delta_{2n,s} ,
\\
\begin{picture}(65,40)
    \put(0,0){\includegraphics[width=2cm]{sunrise.eps}}
    \put(0,5){$s$}
    \put(50,5){$0$} 
\end{picture}
     &=&
  \sqrt{2}
   \sum_{n=1}^\infty
    \left\{
  B_1 (m_{An}, m_{\psi n}) 
  + B_2( m_{An}, m_{\psi n}) 
 \right\} P_L \delta_{2n,s} ,
\\
\begin{picture}(65,40)
    \put(0,0){\includegraphics[width=2cm]{sunrise.eps}}
    \put(0,5){$j$}
    \put(50,5){$s$} 
\end{picture}
    &=&
       \left\{
       B_1 (0,m_{\psi s}) +
       B_2 (0, m_{\psi s}) \right\} \delta_{j s}
\nonumber
\\
  &&
   + B_1 (m_{A j}, 0) i\gamma^5 \delta_{j s}
\nonumber
\\
  &&
   + \sum_{n=1}^\infty
       \left\{
         B_1 (m_{A n} , m_{\psi , j+n})
         +B_2 (m_{A n}, m_{\psi, j+n}) \right\}
         \delta_{j+2n, s}
\nonumber
\\
  &&
   + \sum_{n=1}^\infty
      \left\{
        B_1 (m_{A n} , m_{\psi ,n+s})
         +B_2 (m_{A n}, m_{\psi , n+s})
         \right\} \delta_{j, s+2n}
\nonumber
\\
  &&
   + \sum_{\ell=1}^\infty
     \left\{
        B_1 (m_{A, \ell+s} ,m_{\psi \ell})
        +B_2 (m_{A, \ell+s}, m_{\psi \ell}) \right\}
        i\gamma^5 \delta_{j,s+2\ell}
\nonumber
\\
  &&
   + \sum_{\ell=1}^\infty
     \left\{
       B_1 (m_{A, j+\ell}, m_{\psi \ell})
       -B_2 (m_{A, j+\ell}, m_{\psi \ell})\right\}
       i\gamma^5 \delta_{j+2\ell,s}
\nonumber
\\
  &&
   + \sum_{n=-\infty}^\infty
      \left\{
        B_1 ( m_{An}, m_{\psi ,j+n})
        +B_2 (m_{An}, m_{\psi ,j+n}) \right\} 
        \delta_{j s} ,
\\
\begin{picture}(65,40)
    \put(0,0){\includegraphics[width=2cm]{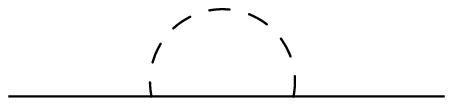}}
    \put(0,5){$0$}
    \put(50,5){$0$} 
\end{picture}
 &=& {\cal N}^2 \left[ \sum_{n=-\infty}^\infty
    E_1 (m_{An}, m_{\psi n}) P_L
     - E_1 (0,0) P_L \right] ,
\\
\begin{picture}(65,40)
    \put(0,0){\includegraphics[width=2cm]{moonrise.eps}}
    \put(0,5){$0$}
    \put(50,5){$s$} 
\end{picture}
   &=&
     -\sqrt{2} {\cal N}^2
       \sum_{n=1}^\infty \left\{
         E_1 (m_{A n}, m_{\psi n}) P_L 
       +
         E_2 (m_{An}, m_{\psi n}) P_R 
        \right\}   \delta_{2n,s} ,
\\
\begin{picture}(65,40)
    \put(0,0){\includegraphics[width=2cm]{moonrise.eps}}
    \put(0,5){$s$}
    \put(50,5){$0$} 
\end{picture}
   &=&
     - \sqrt{2} {\cal N}^2
       \sum_{n=1}^\infty
        \left\{
         E_1 (m_{A n}, m_{\psi n}) 
      +
         E_2 (m_{An}, m_{\psi n}) 
    \right\} P_L 
           \delta_{2n,s} ,
\\
\begin{picture}(65,40)
    \put(0,0){\includegraphics[width=2cm]{moonrise.eps}}
    \put(0,5){$j$}
    \put(50,5){$s$} 
\end{picture}
  &=& {\cal N}^2 \bigg[ i\gamma^5
    E_1 (m_{A j}, 0) \delta_{j s}
\nonumber
\\
  &&
   - \sum_{n=1}^\infty
      \left\{
        E_1 (m_{A n}, m_{\psi , n+s})
        +E_2 (m_{A n} , m_{\psi , n+s})\right\}
        \delta_{j, s+2n}
\nonumber
\\
  &&
   - \sum_{n=1}^\infty
      \left\{
        E_1 (m_{A n} , m_{\psi , j+n})
        +E_2 (m_{A n} , m_{\psi ,j+n}) \right\}
        \delta_{j+2n,s}
\nonumber
\\
  &&
     + \sum_{\ell=1}^\infty
         i\gamma^5
         \left\{
         E_1 (m_{A,j+\ell}, m_{\psi \ell})
         +E_2(m_{A,j+\ell}, m_{\psi \ell})\right\}
         \delta_{j+2\ell,s}
\nonumber
\\
  &&
   + \sum_{\ell=1}^\infty
      i\gamma^5
      \left\{
       E_1 (m_{A, \ell +s}
        , m_{\psi \ell})
        -E_2 (m_{A ,\ell+s}, m_{\psi \ell}) \right\}
        \delta_{j,s+2\ell}
\nonumber
\\
 &&
   + \sum_{n=-\infty}^\infty
      \left\{
       E_1 (m_{A n} , m_{\psi, j+n})
       +E_2 (m_{A n}, m_{\psi, j+n})
\right\} \delta_{j s}
\nonumber
\\
  &&
  - \left\{
  E_1 (0,m_{\psi j}) +E_2(0,m_{\psi j})\right\}
  \delta_{j s} 
  \bigg].
\eea
For $(E_1+E_2)$, the mode sum with 
$-\infty \leq n \leq \infty$ is
regarded as a formal equation because 
$A_{y n}$ has no zero mode.

\subsection{Evaluation of momentum integrals}

We calculate the momentum integrals
by introducing Feynman parameters and
employing the dimensional regularization and 
the Poisson resummation.

The momentum integrals for
$\Pi_{\mu\nu}$ are given by
\bea
  &&
  \sum_{n=-\infty}^\infty
   \left( I_1 (\mu, \nu ; m_{\psi n} , m_{\psi, n+s})
  +I_2 (\mu, \nu; m_{\psi n} ,m_{\psi, n+s})\right) 
\nonumber
\\
  &\!\!\!=\!\!\!&
   {8 i g^2 \over (4\pi)^2 (1+k)} \int_0^1 dx 
   \left\{
     \left( z_4 - \sum_{n_p=1}^\infty
      z_3 e^{-{2 z_4\over z_3}} \cdot \cos 
  (2\pi n_p x s) \right)
 \right.
\nonumber
\\
   && \times  x(1-x)
     \left((p_2^2 -m_{\psi s}^2)\eta_{\mu\nu} 
     -p_{2\mu} p_{2\nu}\right)
\nonumber
\\
  &&
  -\left. {1\over 4}
  \sum_{n_p=1}^\infty
     z_3  (z_3 + 2 z_4)
       e^{-{2 z_4\over z_3}}
    (1-2 x) m_{\psi s} \sin (2\pi n_p x s)
    \eta_{\mu\nu}\right\} .
\eea
For $s=0$, the four-dimensional 
Ward identity is satisfied.
It is also seen from the following equation,
\bea
 &&  I_1 (\mu,\nu; m_1 , m_2)
   + I_2 (\mu, \nu; m_1 , m_2)
 =
 - {8 g^2 \over L}
   \int_0^1 d x \int {d^d \ell\over (2\pi)^d}
   \, {1 \over
   \left[ \ell^2 - \Delta\right]^2 }
\nonumber
\\
   && \times
     \left[
     x(1-x) (p_2^2 \eta_{\mu\nu} -p_{2\mu}p_{2\nu})
    +{1\over 2} (m_1 -m_2) ( x m_2 -(1-x) m_1)
   \eta_{\mu\nu}
   \right] ,
\eea
where
$\ell = p_1 + x p_2$
and 
$\Delta = xm_2^2 + (1-x)m_2^2
-x(1-x) p_2^2$.
In the main text,
the letter of
the external momentum is denoted as $p$
instead of $p_2$.
The momentum integrals for 
$\Pi_{\mu 5}$ are given by
\bea
   &&
   \sum_{n=-\infty}^\infty
    \left( I_1 (\mu; m_{\psi n} , m_{\psi, n+s})
   +I_2 (\mu ; m_{\psi n} , m_{\psi, n+s} )\right)
\nonumber
\\
  &\!\!\!=\!\!\!&
   -{8 i g^2 \over (4\pi)^2 (1+k)}
   p_{2\mu} \int_0^1 d x
    \left\{
    \left( z_4 -\sum_{n_p=1}^\infty
  z_3 e^{-{2 z_4 \over z_3}} \cos (2\pi n_p x s) \right)
     x(1-x) m_{\psi s}
 \right.
\nonumber
\\
  && \left.
  +{1\over 4}  \sum_{n_p=1}^\infty
     z_3 (z_3 +2 z_4)
  e^{-{2 z_4 \over z_3}} (1-2 x) \sin (2\pi n_p xs)\right\} .
\eea
The momentum integrals for $\Pi_{55}$ are
given by
\bea
   && \sum_{n=-\infty}^\infty
     \left( I_1 (m_{\psi n} , m_{\psi, n+s})
      +I_2 (m_{\psi n} , m_{\psi , n+s})\right)
\nonumber
\\
  &\!\!\!=\!\!\!&
  {8 i g^2 \over (4\pi)^2 (1+k)}
   \int_0^1 d x
   \left\{
      z_4 x (1-x) p_2^2  \right.
\nonumber
\\
  &&
   -{1\over 4} \sum_{n_p=1}^\infty
     \left[
      3 z_3^2 (z_3 +2 z_4) + 2 z_3 (2 x (1-x)m_{\psi s}^2)
  \right]
   e^{-{2 z_4 \over z_3}} 
   \cos (2\pi n_p x s) 
\nonumber
\\
  && \left.
   +{1\over 4} \sum_{n_p=1}^\infty
      z_3 (z_3 +2 z_4) 
      e^{-{2 z_4 \over z_3}}
  (1-2 x) m_{\psi s}
    \sin (2\pi n_p x s) \right\} .
\eea

For fermion self-energies,
the momentum integrals with 
$-\infty \leq n \leq \infty$ are given by
\bea
  &&
    \sum_{n=-\infty}^\infty
    \left( B_1 (m_{An}, m_{\psi, n+s})
      +B_2 (m_{An}, m_{\psi, n+s})\right)
      =-{4 i g^2 \over (4\pi)^2}
        \int_0^1 d x {1\over \sqrt{w}}
\nonumber
\\
    &&
  \times
  \left\{ \left(
    w_2 
  -\sum_{n_p=1}^\infty
    w_1 
     e^{-{2 w_2 \over w_1}}
        \cos \left(2\pi n_p {(1+k)^2 x s \over w}\right)
        \right)
  \right.
  (1-x)        
  \left( p\!\!\!/{}_1 -{2 (1+\lambda) m_{\psi s}\over w}\right)
\nonumber
\\
  &&
   \left.
   +\sum_{n_p =1}^\infty
    {(1+k)\over \sqrt{w}}
   w_1 (w_1 +2 w_2) 
    e^{-{2 w_2\over w_1}}
    \sin \left(2\pi n_p
    {(1+k)^2 x s \over w}\right) \right\} .
\eea
Here
\bea
  w_1 &\!\!\!\equiv  \!\!\!&
  {\sqrt{w}\over n_p L} ,\qquad
  w_2 \equiv \sqrt{x(1-x)\left( {(1+\lambda)\over w} 
 m_{\psi s}^2 -p_1^2\right) } ,
\\
 w &\!\!\!\equiv \!\!\!&
  x(1+k)^2 +(1-x) (1+\lambda) .
\eea
The momentum integrals for $E_1$ and $E_2$ 
are obtained as
with the relations
\bea
  E_1 (m_1 , m_2) =
    -{1\over 2} B_1(m_1, m_2) ,
 \qquad
  E_2 (m_1, m_2) =
     {1\over 4} B_2 (m_1 , m_2) .
\eea

\end{appendix}

\newpage



\end{document}